## Magnetodynamics

# All-angle collimation for spin waves

J. W. Kłos, P. Gruszecki, A. E. Serebryannikov, M. Krawczyk

*Faculty of Physics, Adam Mickiewicz University in Poznań, Umultowska 85, 61-614 Poznań, Poland*

*Abstract*— We studied the effect of collimation for monochromatic beams of spin waves, resulting from the refraction at the interface separating two magnetic half-planes. The collimation was observed in broad range of the angles of incidence for homogenous Co and Py half-planes, due to significant intrinsic anisotropy of spin wave propagation in these materials. The effect exists for the sample saturated by in-plane magnetic field tangential to the interface. The collimation for all possible angles of incidence was found in the system where the incident spin wave is refracted on the interface between homogeneous and periodically patterned layers of YIG. The refraction was investigated by the analysis of isofrequency dispersion contours of both pairs materials, i.e., uniform YIG/patterned YIG and Co/Py, which are calculated with the aid of the plane wave method. Besides, the refraction in Co/Py system was studied using micromagnetic simulations.

## I. INTRODUCTION

The propagation of spin waves (SWs) in ferromagnetic films, with the saturated magnetic field applied in the plane of the film, is highly anisotropic. The dispersion relation (frequency vs. wave vector) for SWs varies with changes of direction of the wave vector with respect to the external magnetic field [Stancil 2009]. Such intrinsic anisotropy is typical for magnonic systems [Gieniusz 2013]. The anisotropy of propagation can be also introduced by periodic patterning of the ferromagnetic film [Tacchi 2015]. This leads to the breaking of the symmetry of the dispersion relation even for the configuration which has initially isotropic dispersion, i.e. for a homogeneous film with out-of-plane oriented magnetic field that is known as forward configuration.

The refraction effects on the interfaces between anisotropic media are complex and sometimes counterintuitive at specific conditions [Foteinopoulou 2003, Gralak 2000]. Such effects were intensively explored in photonic systems where the most common way to introduce the anisotropy is patterning of the structure in the form of a photonic crystal [Joannopoulos 2008]. Here we adopt the ideas from photonics that are related to beam refraction [Luo 2002, Chigrin 2004] and collimation [Wu 2012] to magnonic systems, where the intrinsic anisotropy of solid film (absent in photonic systems) can also be exploited. The goal of this paper is to study possibility of collimation for SWs propagating in thin ferromagnetic films, which is, so far, not completely utilized. In magnonics, the control of SWs propagation direction has been discussed in connection with the caustic effect [Veerakumar 2006, Schneider 2010] where the *two* preferred directions of propagation exist. This is related to hyperbolic dispersion of in-plane magnetized thin films. We are interested in collimation, when *only* one direction of propagation is supported. To achieve this effect we need a system characterized by dispersion relation which has flat isofrequency dispersion contour perpendicular to the required direction of the wave propagation. This can be gained in periodic system with rectangular lattice with strong and week coupling in two orthogonal directions [Kumar 2015]. In our study, we obtain suitable dispersion by applying magnetic field in the plane of the system. It makes the propagation of spin waves in orthogonal directions distinguishable (due to introduction of anisotropic dipolar interactions). Moreover, we show that rectangular lattice is not necessary and collimation can be obtained for geometries of higher symmetry.

We consider two cases when collimation can appear: i) due to peculiar dispersion relation of homogeneous ferromagnetic film, and ii) due to tailored magnetic properties of the ferromagnetic film by its regular patterning. In the patterned medium the dispersion relation of SWs can be shaped by adjusting structural parameters. We show, that the periodicity of the patterned medium enables all-angle collimation, i.e., beams incident on the interface between the homogeneous and the periodic medium may formally propagate along the normal direction in the exit medium even of the incidence algle is close to 90° limit. We perform the analysis of the refraction with the aid of isofrequency dispersion contours (IFCs) [Gralak 2003, Jiang 2013]. For the flat interface, the tangential component of the wave vector and, hence, the tangential component of the phase velocity must be conserved. Using this condition, the direction of the group velocity of the refracted wave can be found. It is determined by the direction

Corresponding author: J. W. Kłos(klos@amu.edu.pl).

normal to the IFC at the point on this contour where the tangential component of the phase velocity is the same as in the medium from which the wave is incoming. In turn, the directions of incidence and refracted beams are given by the directions of the corresponding group velocities. Therefore, the direction of the beam refraction can be deduced by the geometrical analysis of the IFCs shape for two homogeneous media at the opposite sides of the flat interface. Similar analysis can be performed for the wave incident on the interface between homogeneous and periodic media. However, in this case the wave vector can be shifted by any reciprocal lattice vector's tangential to the interface due to discrete translational symmetry along the interface. We've used this approach to adjust the geometry of the system and find the IFCs suitable for obtaining the all-angle collimation in yttrium iron garnet (YIG) based antidots lattice.

In this letter, we will discuss the geometry of planar magnonic systems with a single internal interface. Then, we consider the SW dispersion and micromagnetic simulations (MSs) results for a system comprising only homogeneous films. Finally, we present a detailed analysis of dispersion for a system containing a square-lattice magnonic crystal. The paper will be finished with short conclusion.

## II. THE MODEL

We consider two planar systems that consist of two half-planes. The first system is composed of two uniform half-planes of Co and Py films (see Fig.1b). For this system the collimation is limited to some range of incidence angels of SWs propagating from the side of Co. In turn, the all-angle collimation is showed for the structure made of the YIG with one half-plane in the form of a homogeneous film and the second half-plane in the form of a square lattice of cylindrical antidots (see Fig.2a and Fig.3c). The interface with solid YIG is along one of the principal directions of the square lattice, i.e., is parallel to the $x$-axis. The magnetic field $H_0$ is applied along the interface for both systems.

To describe the SW dynamics, we used Landau-Lifshitz equation (LLE) with exchange and dipolar interactions included. The two-dimensional dispersion relations (for both homogeneous and patterned media) were calculated using plane wave method (PWM) for the saturation state of the static magnetization [Rychly 2015]. The simulations of refraction and reflection of the Gaussian beams were performed with the aid of MSs using Mumax3 [Vansteenkiste 2014] package solving the full LLE based on finite-difference time-domain method. Detailed algorithm of MSs and especially description of Gaussian beams of SWs generation in MSs is presented in [Gruszecki 2015].

## III. THE RESULTS AND DISCUSSION

The collimation effect can be observed for a magnetic planar system without periodic patterning. Fig.1 presents the simulated refraction of the Gaussian beam of SWs on the interface between Co and Py. Both materials, and especially Co, are anisotropic in terms of SW propagation. The other factor which increases the anisotropy of Co (in reference to Py) is the fact that the selected frequency, 20 GHz, is much closer to the FMR frequency of Co than to that of Py, due to higher magnetization saturation. As a result we observe the highly anisotropic IFC for Co (drawn in red in Fig. 1a), exhibiting hyperbolic sections. The IFC for Py has stadium-like shape. To observe the collimation of SWs incident from Co after refraction at the interface with Py, we use the nearly flat sections of the stadium-like IFC (drawn in blue in Fig. 1a). However, the observed collimation is restricted here to some range of incidence angles around the direction normal to the interface, i.e., for the SWs with small tangential component of the phase velocity. This condition means that we are using the hyperbolic section of the IFC in Co in order to obtain collimation in Py. The limited collimation incidence angle is close to the parabolic point [Chigrin 2004] of IFCs, which delimits the areas of the hyperbolic dispersion.

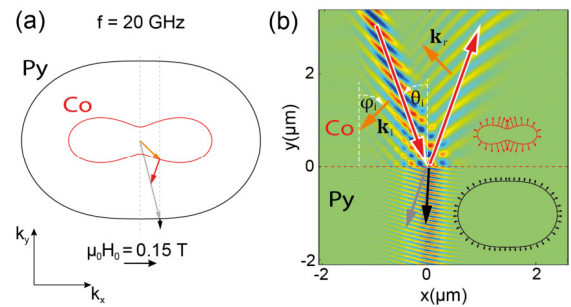

Fig. 1. (color online) Refraction of the SWs with frequency 20 GHz on the interface between Co and Py films (a) from dispersion results obtained by using PWM and (b) from MSs where the amplitude of the out-of-plane component of magnetization was visualized. The thickness of the films is 12 nm; the external magnetic field (0.15 T) is applied along the interface. Directions of the incident and reflected beams are marked by red arrows, direction of the refracted beam is marked by black arrow, for Co and Py respectively. The directions of phase velocity are marked by orange and gray arrows, for Co and Py, respectively. $\theta_i$ and $\varphi_i$ denote the angle of incidence and the angle spanned between wave-vector $\mathbf{k}_i$ and normal to the interface. Note the difference in divergence of the outgoing beams in Co and Py.

Fig. 1b demonstrates refraction and reflection of the SW's Gaussian beam at the interface between the Co and Py thin films that are simulated with the aid of Mumax3 package. Both films were 12 nm thick, 8 μm wide and 8 μm long (discretized using 8 nm × 8 nm × 12 nm rectangular unit cells), adjusting each other at one straight horizontal interface. The system was assumed to be uniformly and stably magnetized, and simulated under the presence of a small (0.15 T), static, in-plane, directed parallel to the interface, magnetic field. In the Co film, far from the edges, the Gaussian beam of SWs was continuously excited, with the width of 0.5 μm and at the frequency $f$=20 GHz. The wave vector of the incident beam, $\mathbf{k}_i$, creates an angle $\varphi_i$=45° with the normal to the interface. To stay in the linear regime, the amplitude of the excitation was much smaller than the saturation magnetization of Co. In calculations, typical magnetic parameters were assumed: saturation magnetizations $M_{S,Co}=1.45\times10^6$ A/m, $M_{S,Py}=0.7\times10^6$ A/m and exchange constants $A_{ex,Co}=3\times10^{-11}$ J/m, $A_{ex,Py}=1.1\times10^{-11}$ J/m for Co and Py, respectively. Simulations of the SWs propagation were carried out at a finite value of the damping parameter, α=0.0005, for the Co and Py films.

In Co the phase velocity of SWs (marked by orange arrow, being perpendicular to the wave-fronts and parallel to the wave vector $\mathbf{k}_i$) is almost perpendicular to the group velocity (marked by red arrow and denoting the direction of an energy propagation), see Fig.1b. This effect is a result of the hyperbolic-like IFC (red line in Fig.1a) occurring in Co with the chosen parameters. Similar analysis of MSs' results and IFC of Py shows that the angle between group and phase velocities is much smaller, but still significant. As seen in Fig.1b, the group velocity in Py is almost perpendicular to the interface in a range of the angles φ$_i$ up to ~60°. This can be concluded from Fig.1a and was confirmed by multiple MSs for various angles of incidence.

The difference in divergence of the beams propagating in Co and Py is worth to notice. It results from different values of phase velocities for SWs of the same frequency in Co and Py. On both sides of Co/Py interface we have then different ratios of wavelength to width of spin wave beam, i.e. in the case of Co the ratio is greater and due to that divergence of beam is stronger [Saleh 2007]. Nevertheless, in metallic ferromagnets, especially in Co, the damping is high. This significantly limits the proposed collimation scenario for practical realization.

Collimation of SWs can be obtained and the mentioned restriction related to damping can be mitigated with the use of magnonic crystals based on thin YIG film [Yu 2014]. The two dimensional plot of the dispersion relation of the planar square, antidots lattice based on YIG film in Fig.2a is presented in Fig.2b. For the magnetic field applied in-plane, the dispersion is strongly anisotropic for lower magnonic bands [Kłos 2012], where dipolar interactions prevail over the exchange ones.

The magnetic field applied in the $x$-direction supports the propagation of SWs (non-zero slope of the dispersion relation) along the $y$-direction in low frequency range. This is mostly caused by the profile of demagnetizing field lowering the effective magnetic field in the rows between antidots [Tacchi 2015] which are aligned in the $y$-direction. It is also clearly visible in Fig.2b, where the narrower frequency gaps and wider bands resulting in larger group velocities are observed along this direction. The top of the first band and the bottom of the second band have a peculiar shape, i.e., the dispersion is almost independent on the $k_x$ component of the wave vector. This means that group velocity, being the gradient of the dispersion relation, is directed almost exactly along the $y$-direction, independently on the direction of the corresponding phase velocity. The Bloch waves of different $\mathbf{k}$-vectors will then propagate mostly in the $y$-direction, enabling collimation.

To investigate this effect, we've considered the refraction of the plane waves entering the patterned YIG film from the homogeneous region of this material. We have chosen arbitrarily chosen on of the frequencies of 9.1 GHz, laying within the second band in the center of the region where the frequency of SWs is almost independent on $k_x$ and varies almost linearly with $k_y$ (see white dashed line in Fig.2b). The top of the first band could also be considered to gain this effect but this band is much narrower and characterized by lower group velocities and much narrower range of linear decadence of the frequency on $k_y$ component of the wave vector. This basically limits its application to collimation of monochromatic beams only.

Fig.3a presents the IFCs for the homogeneous film of YIG (red line) and for the patterned one (black lines, extracted from the magnonic band structure shown in Fig.2b). The IFC for the homogeneous layer of YIG is approximately circular because of the weak intrinsic anisotropy. It is caused by small magnetization saturation of YIG and a comparatively high value of the considered frequency, which is significantly above the FMR frequency. Therefore the phase velocity and group velocity are almost *collinear* for the SWs at this and higher frequencies.

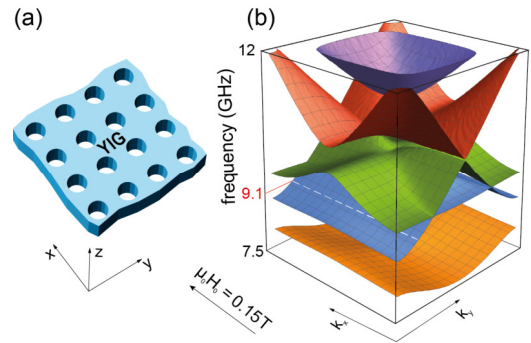

Fig. 2. (color online) (a) The structure of the magnonic antidots lattice in the form of YIG film of thickness 12 nm patterned by cylindrical antidots (diameter 84.6 nm) arranged in square lattice (lattice constant 150 nm). The values of magnetic parameters for YIG are: $M_S$=0.194x10$^6$ A/m, $A_{ex,Co}$=0.40×10$^{-11}$ J/m. (b) Dispersion relation for the considered structure with external field applied along one principal direction of antidots lattice (i.e. $x$-direction), calculated using PWM. The figure presents four lowest (and the part of fifth) magnonic bands over the whole first Brillouin zone. The white dashed line in (b) marks the IFC of the SW at 9.1 GHz.

The IFCs at 9.1GHz for the antidots lattice in Fig.3a are approximately straight and stretch between the boundaries of the first Brillouin zone. Due to periodicity of the dispersion relation in dependence on the wave vector, the IFCs are extended infinitely in the reciprocal space, without any gaps. Also the group velocity (marked by black arrows) almost coincides with one of two opposite defections *(y* or *–y)* and has a nearly constant value at any point of the IFC. This peculiar shape of the IFC is responsible for all-angle collimation. Indeed, there is no restriction regarding the tangential component of the phase velocity that would result from the dispersion of the patterned YIG and make obtaining of this effect impossible. Therefore, the SWs incoming form the homogeneous YIG propagate in the direction normal to the interface, independently on their incidence angle. It is illustrated in Fig.3b, where IFCs are shown together with directions of the phase and group velocities in the media on the both sides of the virtual interface, for three selected incidence angles. One more illustration of the collimation effect is schematically presented in Fig.3c, where the directions of incident and refracted waves are imposed on the sketch of the considered structure. The studies based on geometrical analysis of the shape of IFCs can be supplemented by the MSs results for refraction of Gaussian beams, as it was done above for the interface of two homogenous materials, i.e. Co and Py.

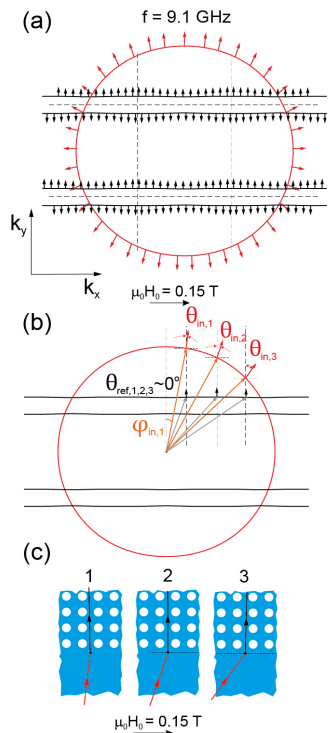

Fig. 3. (color online) (a) The IFCs for the antidots lattice presented in Fig. 2a and solid thin film of YIG at the frequency 9.1 GHz are shown in the wave vector plane with black and red solid lines, respectively. Arrows show directions of the group velocity at different values of $k_x$. (b), (c) Schematics of all-angle collimation effect arising due to the flat IFCs for antidots lattice; beams coming from the solid film side are collimated in the patterned YIG film with antidots lattice into one beam propagating along the normal to the virtual interface ($\theta_{ref}=0$), independently on the incidence angle $\theta_{in}$. The angle $\varphi_{in}$ corresponds to the angle spanned between exemplary wave vector (or phase velocity) of incident SWs and normal to the interface.

## IV. CONCLUSION

We have demonstrated the effect of all-angle collimation for SWs in YIG based square-lattice magnonic crystal. It is shown that for the SWs incidents from the homogeneous region of YIG at an arbitrary angle, the outgoing wave propagates in the properly patterned area of the same material in the normal direction. This enables collimation without any restriction, except for that related to the coupling efficiency at the virtual interface. We supplemented this investigation by the analysis of SW refraction on the interface between two homogeneous materials: Co and Py. It was found in this structure, that the collimation effect can exist in a limited range of the incidence angle variation, being caused by intrinsic anisotropy of magnetic films.

## ACKNOWLEDGMENT

The funding from Polish National Science Centre (DEC-2-12/07/E/ST3/00538) and EU's Horizon2020 programme under the Marie Skłodowska-Curie grant No644348 is acknowledged.

## REFERENCES

Chigrin D N (2004), "Radiation pattern of a classical dipole in a photonic crystal: Photon focusing," Phys. Rev. E 70, 056611, doi: 10.1103/PhysRevE.70.056611.

Foteinopoulou S, Soukoulis C M (2003), "Electromagnetic wave propagation in two-dimensional photonic crystals: A study of anomalous refractive effects," Phys. Rev. A 79, 033829, doi: 10.1103/PhysRevB.72.165112.

Gieniusz R, Ulrichs H, Bessonov V D, et al. (2013)," Single antidot as a passive way to create caustic spin-wave beams in yttrium iron garnet films," Appl. Phys. Lett. 102, 102409, doi:10.1063/1.4795293.

Gralak B, Enoch S, and Tayeb G (2006), Superprism effects and EGB antenna applications" chap.10 in "Metamaterials - Physics and Engineering Explorations," Eds. Engheta N, Ziolkowski W R, John Wiley & Sons, pp. 261-284.

Gruszecki P, Dadoenkova Yu S, Dadoenkova N N, et. al (2015),"Influence of magnetic surface anisotropy on spin wave reflection from the edge of ferromagnetic film," Phys. Rev B 92,054427, doi:10.1103/PhysRevB.92.054427.

Gralak B, Enoch S, and Tayeb G (2000), "Anomalous refractive properties of photonic crystals," J. Opt. Soc. Am. A 17, 1012.

Jiang B, Zhang Y, Wang Y, et. al (2012), "Equi-frequency contour of photonic crystals with the extended Dirichlet-to-Neumann wave vectoreigenvalue equation method," J. Phys. D: Appl. Phys. 45, 065304, doi:10.1088/0022-3727/45/6/065304.

Joannopoulos J D, Johnson S G, Winn J N, et. al (2008), "Photonic Crystals:Molding the Flow of Light," Princeton University Press.

Kumar D, Adeyeye A O (2015), "Broadband and total autocollimation of spin waves using planar magnonic crystals," J. Appl. Phys. 117, 143901, doi: 10.1063/1.4917053.

Kłos J W, Sokolovskyy M L, Mamica S, et. al (2012), "The impact of the lattice symmetry and the inclusion shape on the spectrum of 2D magnonic crystals," J. Appl. Phys. 111, 123910, doi:10.1063/1.4729559.

Luo C, Johnson S G, Joannopoulos J D, and Pendry J B (2002), "All-angle negative refraction without negative effective index," Phys. Rev. B 65,201104, doi: 10.1103/PhysRevB.65.201104.

Rychły J, Gruszecki P, Mruczkiewicz M, et. al (2015), "Magnonic crystals — prospective structures for shaping spin waves in nanoscale," Low Temp. Phys. 41, 959.

Saleh B E A, Teich M C (2007), "Beam optics" chap.3 in "Fundamentals of Photonics, 2nd Edition," Eds. Saleh B E A, Teich M C, John Wiley & Sons, pp. 75-85.

Schneider T, Serga A A, Chumak A V, et. al, "Nondiffractive Subwavelength Wave Beams in a Medium with Externally Controlled Anisotropy," (2010) Phys. Rev. Lett. 104, 197203, doi: 10.1103/PhysRevLett.104.197203.

Stancil D, Prabhakar A (2009), "*Spin waves – theory and applications*," Springer.

Tacchi S, Gruszecki P, Madami M, el. al (2015), "Universal dependence of the spin wave band structure on the geometrical characteristics of two-dimensional magnonic crystals," Sci. Rep. 5, 10367, doi:10.1038/srep10367.

Vansteenkiste A, Leliaert J, Dvornik N, et al. (2014), "The design and verification of MuMax3," AIP Adv. 4, 107133, doi: 10.1063/1.4899186.

Veerakumar V and Camley R E (2006), "Focusing of Spin Waves in YIG Thin Films," IEEE Trans. Mag. 42, 3318, doi: 10.1109/TMAG.2006.879624

Wu Z H, Xie K, Yang H J, et. al (2012), "All-angle self-collimation in two-dimensional rhombic-lattice photonic crystals," J. Opt. A 14, 15002, doi:10.1088/2040-8978/14/1/015002.

Yu H, Kelly O A, Cros V, et. al (2014) "Magnetic thin-film insulator with ultra-low spin wave damping for coherent nanomagnonics," Sci. Rep. 4, 6848, doi: 10.1038/srep06848.